\documentclass[aps,prb,twocolumn,prabib,amsmath,amssymb,superscriptaddress,notitlepage]{revtex4-1}
\usepackage{graphics}
\usepackage{bm}
\usepackage{dcolumn}
\usepackage{natbib}
\usepackage{epstopdf}
\usepackage{float}
\usepackage{graphicx}
\usepackage{epsfig}
\usepackage[pdfstartview=FitH]{hyperref}
\usepackage{color}
\usepackage{appendix}
\usepackage{ulem}

\begin{document}
\title{Prethermal time crystals in a one-dimensional periodically driven Floquet system}
\author{Tian-Sheng Zeng}
\affiliation{Department of Physics and Astronomy, California State University, Northridge, California 91330, USA}
\author{D. N. Sheng}
\affiliation{Department of Physics and Astronomy, California State University, Northridge, California 91330, USA}
\date{\today}
\begin{abstract}
Motivated by experimental observations of time-symmetry breaking behavior in a periodically driven (Floquet) system, we study a one-dimensional spin model to explore the stability of such Floquet discrete time crystals (DTCs) under the interplay between interaction and the microwave driving. For intermediate interactions and high drivings, from the time evolution of both stroboscopic spin polarization and mutual information between two ends, we show that Floquet DTCs can exist in a prethermal time regime without the tuning of strong disorder. For much weak interactions the system is a symmetry-unbroken phase, while for strong interactions it gives its way to a thermal phase. Through analyzing the entanglement dynamics, we show that large driving fields protect the prethermal DTCs from many-body localization and thermalization. Our results suggest that by increasing the spin interaction, one can drive the experimental system into optimal regime for observing a robust prethermal DTC phase.
\end{abstract}
\maketitle
\section{Introduction}
Recently, Wilczek proposed a concept of quantum time crystals which spontaneously break the continuous time translational symmetry into a discrete case~\cite{Wilczek2012}. However for thermodynamic equilibrium system, this kind of time crystal could not exist following a no-go theorem~\cite{Bruno2013,Watanabe2015}. Alternately, Sacha first proposed a new strategy using a periodically driven (Floquet) system to simulate the discrete time-translation symmetry breaking~\cite{Sacha2015}. Later, Khemani {\it et.~al.} in Ref.~\cite{Khemani2016} and Else {\it et.~al.} in Ref.~\cite{Else2016} respectively construct a more concrete example of time symmetry breaking in non-equilibrium Floquet eigenstates of matter without stationary analogs~\cite{Moessner2017}. In the presence of high-frequency driving and strong disorder, the Floquet eigenstates realize Floquet-many-body-localization (Floquet-MBL), which are robust against the eigenstate thermalization~\cite{Abanin2016,Ponte2015,Lazarides2015} and can host symmetry-broken or topological-ordered phases in highly excited states~\cite{Bauer2013,Huse2013,Bahri2015,Keyserlingk2016a,Keyserlingk2016b}. Such a phase that spontaneously breaks discrete time-translational symmetry is dubbed the Floquet ``discrete time crystals'' (DTC), which have been demonstrated in recent theoretical and numerical studies~\cite{Else2016,Keyserlingk2016,Yao2017,Lazarides2017,Sacha2017}.

In pioneering experiments, the DTCs are observed in one-dimensional trapped ions~\cite{Zhang2017} and three-dimensional spins in diamond~\cite{Choi2017} with long range dipolar interaction and random disorder.
However, the disorder strength in Ref.~\cite{Choi2017}, is still much weaker than what is desired for driving an interacting three-dimensional system into a MBL phase, especially in the presence of long range dipolar interaction for the experimental system, which may further enhance delocalization in three dimensions~\cite{Burin2015}. The disorder strength is also much weaker than the microwave driving field, which makes the connection between the observed DTC and the disorder-protected Floquet-MBL ambiguous. Thus, the experimentally observed DTC demands further theoretical studies. A perturbative study of this 3D related model without disorder leads to a critical time crystal~\cite{Ho2017}. For general periodically driven systems, when subject to a high-frequency periodic driving, the energy absorption rate is exponentially small in the driving frequency for any initial state~\cite{Abanin2015,Mori2016}, and it is shown that Floquet time crystals can be stabilized in the so-called prethermal phases without the need for strong disorder, before eventually heating into a thermal phase~\cite{Else2017,Abanin2017}. Another possibility is that the observed DTC falls into the short time regime before reaching the prethermal state, as suggested in Ref.~\cite{Else2017}.
Clearly, by tuning driving frequency and other couplings, the complex interplay between interactions and disorder may either further strengthen DTC behavior or lead to competing phases~\cite{Kucsko2016}.
Recently, possible realizations of clean Floquet DTC are proposed in different non-localized setups~\cite{Huang2017,Russomanno2017}, but not in a prethermal regime. Nevertheless, the existence of Floquet DTC without disorder indicates other possible physical mechanisms instead of the Floquet-MBL nature, and would demand a more careful theoretical explanation of the DTC behavior in periodically driven systems, which also stimulates us to investigate the role of disorder in our theoretical models.
Taken together, these theoretical and experimental studies  have raised  critical issues regarding the nature of possible non-equilibrium quantum phases
in such systems and the mechanism for the experimentally observed DTC behaviors remains not well understood.

In this work, we study the time evolutions  of one-dimensional periodically driven Floquet interacting spin chain  with disorder, and address the possible prethermal DTC in non-localized systems. We establish the existence of the prethermal DTC phase identified by the steady plateaux in the nonequilibrium dynamics of the system (e.g., evolution of any local observables)~\cite{Else2017,Abanin2017} at times $t\lesssim t^{\ast}$, from two signatures: a) steady stroboscopic spin polarization in each Floquet cycle
and b) robust mutual information between two ends. From the dynamics of the spin polarization and mutual information, we map out the phase diagram and illustrate the dynamical phase transitions by tuning interaction and other couplings like spin flipping terms as in experiments.

This paper is organized as follows. In Sec.~\ref{model}, we give a description of the one-dimensional time-dependent periodic spin Hamiltonian, as experimentally realized. In Sec.~\ref{ground}, we present the phase diagram of prethermal DTC from the nonequilibrium dynamics under the interplay among interactions, time scale, disorder strength, and driving field via the time-dependent density matrix renormalization group method. In Sec.~\ref{drivingfield}, we discuss the effect of driving field through analyzing the entanglement dynamics. In Sec.~\ref{dynamicaltransition}, we discuss the dynamical phase transitions by tuning coupling parameters. Finally, in Sec.~\ref{summary}, we summarize our results and discuss the experimental implication.

\section{The Model Hamiltonian}\label{model}
We consider the one-dimensional time-dependent periodic spin Hamiltonian $H(t)=H(t+T)$,
\begin{align}
  &H(t)=\sum_i\Omega_xS_{i}^{x}+\Delta_i S_{i}^{z}+H_{i,i+1},0<t<\tau_1,\label{H1}\\
  &H(t)=\sum_i\Omega_y(1-\epsilon)S_{i}^{y}+\Delta_i S_{i}^{z}+H_{i,i+1},\tau_1<t<T,\label{H2}
\end{align}
with $H_{i,i+1}=J(S_{i}^{x}S_{i+1}^{x}+S_{i}^{y}S_{i+1}^{y}-S_{i}^{z}S_{i+1}^{z})$ and the time period $T=\tau_1+\tau_2$. $\Omega_x (\Omega_y)$ is the Rabi frequency of the microwave driving. $\Delta_i\in(-W,W)$ is a quasirandom disorder~\cite{Schreiber2015}. $J$ is the strength of the orientation dependent nearest-neighbour spin couplings~\cite{note}. Given that $\Omega_x (\Omega_y)\gg W,J$ in experiments, an effective $S^x$-polarization-conserving Hamiltonian during $\tau_1$ is given by
\begin{align}
  H_{e}(t)=\sum_i\Omega_x S_{i}^{x}+J S_{i}^{x}S_{i+1}^{x},\nonumber
\end{align}
while the global spin rotation term
\begin{align}
  H_{e}(t)=\sum_i\Omega_y S_{i}^{y}\nonumber
\end{align}
during $\tau_2$ flips the $x$-direction spin polarization $S^x$, which illustrates the $2T$-periodic response of the system as discussed in Refs.~\cite{Choi2017,Ho2017}.

\begin{figure}[t]
  \includegraphics[height=2.45in,width=3.4in]{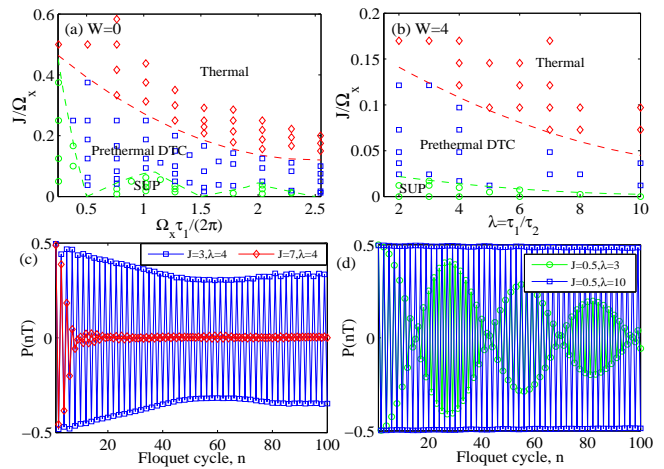}
  \caption{\label{phase}(Color online) Phase diagram plotted in the parameter plane (a) the phase $\Omega_x\tau_1/2\pi$ and the ratio $J/\Omega_x$ at $\lambda=4$ without disorder $W=0$ and (b) the ratio $\lambda$ and the interaction $J/\Omega_x$ at $\Omega_x=41.2$ with disorder $W=4$. The time evolution of spin polarization $P(nT)$ under disorder $W=4$ in Fig.~\ref{phase}(b) for (c) different interactions $J$ at $\lambda=4$ and (d) different interaction time scales at weak interaction $J=0.5$. In the intermediate interaction regime, a DTC behavior can occur in a prethermal regime of the system without the need for fine tuning and strong disorder. The green dashed boundary line is depicted by the emergence of the steady plateaux of $P(nT)$ and $I(nT)$ as in Fig.~\ref{mut}, while the red dashed boundary line indicates the disappearance of the steady plateaux of $P(nT)$ and $I(nT)$ when $J$ increases as in Fig.~\ref{thermal}. The parameters $\epsilon=0.038,\Omega_y\tau_2=\pi$.}
\end{figure}

In close relationship to experiments, typically we take the parameters $\epsilon=0.038,\Omega_y=10\pi,\tau_2=0.1$ unless otherwise specified, thus, $\Omega_y\tau_2=\pi$ takes the role of spin flipping. In the following, we explore the non-equilibrium Floquet dynamics $U(T)=\mathcal{T}\exp(-i\int_{0}^{T}H(t)dt)$ of the full Hamiltonian from Eqs.~\ref{H1} and~\ref{H2} in one-dimensional spin chain, and discuss the effects of interactions $J$, time scale $\lambda=\tau_1/\tau_2$, disorder strength $W$, and driving field $\Omega_x$ via the time-dependent density matrix renormalization group method based on a matrix product state (MPS) representation~\cite{Dolfi2014} with a maximal bond dimension 300, which allows well converged results. For the time evolution, we take the time step $\Delta t=0.01$, and use a second-order Trotter decomposition within this time step, up to a maximum chain length $L=30$.

According to two equivalent definitions of Floquet time crystal~\cite{Else2016,Khemani2016b}, to probe the existence of time-crystalline order, we choose two different kinds of the initial states: (i) the random product  initial states $|\psi_s(t=0)\rangle=|\{s_i\}\rangle$, where $|s\rangle,s=\pm\frac{1}{2}$ are the eigenvectors of the Pauli spin operator $S^x$, 
in order to demonstrate the broken time-translational symmetry from the expectation value of local spin polarization in the middle of the chain $P(t)=\langle\psi(t)|S_{i=L/2}^{x}|\psi(t)\rangle\neq P(t+T)$ at stroboscopic times; (ii) the long-range correlated initial states $|\psi_l(t=0)\rangle=\frac{1}{\sqrt{2}}\left(e^{i\tau_1E_{h}}|\{s_i\}\rangle\pm e^{-i\tau_1E_{h}}|\{-s_i\}\rangle\right)$, in order to demonstrate the broken time-translational symmetry from the mutual information between the two ends of the spin chain $I(t)=S_A+S_B-S_{AB}\sim\log2$, where $S_{A}(S_{B})$ is the von Neumann entropy of the reduced density matrix for site A(B). Here $E_{h}$ satisfies $\sum_i\Omega_x S_{i}^{x}|\{s_i\}\rangle=2E_{h}(\{s_i\})|\{s_i\}\rangle$, and $|\psi_l\rangle$ becomes Floquet eigenstates of the evolution operator $U_e(T)=\mathcal{T}\exp(-i\int_{0}^{T}H_e(t)dt)$ of the effective Floquet Hamiltonian $H_{e}(t)$, which exhibits the spatiotemporal order~\cite{Keyserlingk2016}.

We emphasize that $|\psi_l\rangle$ is not the true Floquet eigenstates of the evolution operator $U(T)$ of the Floquet Hamiltonian $H(t)$, mainly for two reasons: (i) numerically we cannot obtain the exact highly-excited eigenstates of $U(T)$ from exact diagonalization for large system sizes $L>20$ in this non-integrable Hamiltonian model Eq.~\ref{H1}; (ii) using $|\psi_l(t=0)\rangle$ as an approximating solution of $U(T)$, we can explore the robustness of DTC from the mutual information in its non-equilibrium dynamics $|\psi_l(t=nT)\rangle=(U(T))^n|\psi_l(t=0)\rangle$ by considering the perturbation effects of traverse interaction term $J(S_{i}^{y}S_{i+1}^{y}-S_{i}^{z}S_{i+1}^{z})$, time scale $\lambda=\tau_1/\tau_2$, disorder strength $W$, and driving field $\Omega_x$, and study the possible emergence of prethermal Floquet steady states at stroboscopic times.

\begin{figure}[t]
  \includegraphics[height=2.1in,width=3.4in]{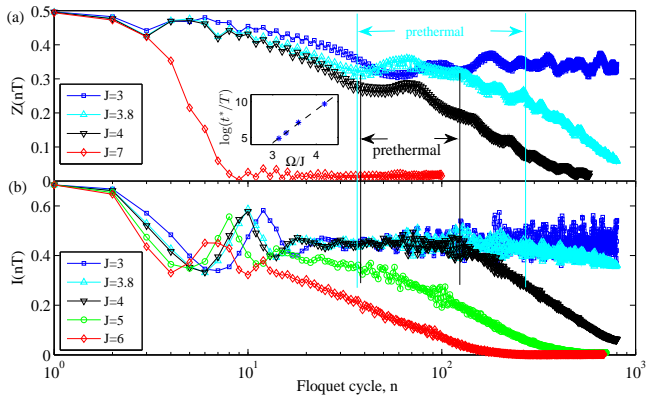}
  \caption{\label{thermal}(Color online) (a) The stroboscopic spin polarization $Z(t=nT)$ and (b) the mutual information $I(nT)$ with the increasing interaction strength $J$. A prethermal DTC phase is indicated by a robust plateau of both $Z(nT)$ and $I(nT)$ at intermediate times $nT\lesssim t^{\ast}$ before eventually heating up to a thermal phase after $nT>t^{\ast}$. The inset shows an exponential growth of $t^{\ast}$ vs $\Omega/J$. The parameters $W=4,\lambda=4,\Omega_x=41.2,\epsilon=0.038,\Omega_y\tau_2=\pi$. }
\end{figure}

\section{Phase Diagram}\label{ground}
Here we study the interplay between local energy scale $\Omega_x$, interaction strength $J$ and interaction time scale $\lambda=\tau_1/\tau_2$. Fig.~\ref{phase}(a) plots the phase diagram in parameters $(\Omega_x,J/\Omega_x)$  without disorder $W=0$, while Fig.~\ref{phase}(b) plots the phase diagram in parameters $(\lambda,J/\Omega_x)$ with a moderate disorder strength $W=4$. For both cases, a prethermal DTC phase can appear in the sandwich region for an intermediate ratio $J/\Omega_x$.

Figs.~\ref{phase}(c) and~\ref{phase}(d) depict the distinguishing oscillating behaviors of $P(nT)$ in the different phase regions of Fig.~\ref{phase}(b) by tuning interaction $J$ and $\lambda$ for fixed driving field and disorder. For very strong interactions $J\gg1$, $P(nT)$ decays rapidly and the $2T$-periodicity is completely destroyed, where the phase is a thermal phase, as indicated in Fig.~\ref{phase}(c). For very weak interactions $J\ll1$, $P(nT)$ exhibits an uncorrelated oscillation with two incommensurate peaks of the corresponding Fourier spectra $S(\omega)=\sum_{n}P(nT)e^{i2\pi n\omega}$ at $\omega=(1\pm\epsilon)/2$, where the phase is taken as a symmetry-unbroken phase (SUP), as indicated in Fig.~\ref{phase}(d). Only for an intermediate interaction strength, spin polarization $P(nT)$ alternates between positive and negative values as a steady plateaux for a sufficiently long time, resulting in a subharmonic peak at $\omega=1/2$, where the phase is identified as a prethermal DTC phase, until the time that is nearly exponentially long in the ratio of interaction strength to the drive frequency when the system finally gives its way to thermalization.

In our model, the dominant non-interacting local energy $\Omega_xS^x\thicksim \Omega_x/2$ is comparable to the drive frequency $\Omega=2\pi/T$, while the local interaction energy $JS_i^xS_{i+1}^x\thicksim J/4$ is much small compared to $\Omega$. In agreement with the analysis of Ref.~\cite{Else2017,Abanin2017}, for $J/\Omega_x,J/\Omega\ll1$, the time evolution of our system would exhibit a prethermal behavior at times $t_{\text{pre}}<t<t^{\ast}$, where $t_{\text{pre}}$ is the short thermalization time scale of order $2\pi/J$, and $t^{\ast}/T\simeq\exp(C\Omega/J)$ the characteristic heating time depending on the ratio of driving frequency $\Omega=2\pi/T$ to local coupling energy scale $J$ with $C$ a numerical coefficient of order 1. Typically $t_{\text{pre}}\ll t^{\ast}$, for interaction parameters $J\sim1\ll\Omega$. For strong interactions comparable to the driving frequency, $t_{\text{pre}}\sim t^{\ast}$, the prethermal regime may be destroyed. Figs.~\ref{thermal}(a) and~\ref{thermal}(b) show the dephasing heating effects as the interaction is increased further. The prethermal DTC phase indicated by a robust plateau of both stroboscopic spin polarization~\cite{Else2016} $Z(nT)=(-1)^nP(nT)$ and the mutual information $I(nT)$ at intermediate times $nT\lesssim t^{\ast}\sim\exp(C\Omega/J)$ which is fitted in the inset of Fig.~\ref{thermal}(a). When $nT\gtrsim t^{\ast}$, both $Z(nT)$ and $I(nT)$ show an exponentially decaying behavior in the thermal regime. By increasing $J$, the heating time $t^{\ast}$ collapses exponentially, and the system becomes a completely thermal phase in the strongly interacting regime once $t_{\text{pre}}\simeq t^{\ast}$.

\begin{figure}[t]
  \includegraphics[height=1.5in,width=3.4in]{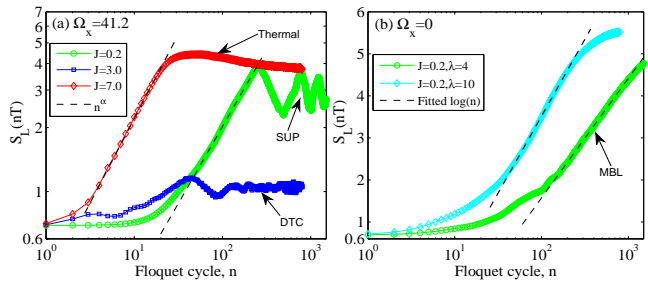}
  \caption{\label{mbl}(Color online) Dynamics of the half-chain entanglement entropy $S_L(t)$ with the initial superposition states $|\psi_l(t=0)\rangle$ for: (a) different interactions $J$ at $\Omega_x=41.2,\lambda=4$ and (b) different interaction time scales $\lambda$ at $\Omega_x=0,J=0.2$, respectively. The parameters $W=4,L=20,\epsilon=0.038,\Omega_y\tau_2=\pi$. }
\end{figure}

\section{Effects of driving field $\Omega_x$}\label{drivingfield}
Having built up the stability of a DTC for the intermediate regime of interaction, we turn to an analysis of the robustness of DTC from the disorder effects. Here one important issue is whether the presence of disorder plays an essential role on the stability of the observed DTC phase. The interplay between interaction and disorder may lead to MBL, where a stable Floquet-MBL phase may occur for weak interaction, strong disorder and high driving frequency~\cite{Abanin2016,Ponte2015,Lazarides2015}. The  hallmark signature of MBL is the unbounded logarithmic growth of the half-chain subsystem
entanglement entropy~\cite{Ponte2015,Bardarson2012,Torre2016}, which can be used to distinguish the MBL phase from other phases.

For large driving field $\Omega_x\gg W$, with the initial superposition state $|\psi_l(t=0)\rangle$, we plot its entanglement dynamics in Fig.~\ref{mbl}(a) in the different phase regions of Fig.~\ref{phase}(b). In the early time regime $n\lesssim10$, the entanglement grows slowly. In the later regime $n>10$, the entanglement entropy $S_L(nT)$ increases as $n^{\alpha}$ against the Floquet cycle, as indicated by the fitted dashed line in Fig.~\ref{mbl}(a). This power law behavior of ballistic spreading of entanglement is shown to be general in a diffusive nonintegrable system~\cite{Kim2013,Zhou2017} observed here for both SUP and thermal phases. In a longer time regime $n>10^2$ for weak interaction $J\ll W$, the entanglement entropy of the SUP oscillates around a large value with large entropy fluctuations, and converges slowly, which is expected for the effective dominating integrable Ising model $H_{e}(t)=\sum_i\Omega_x S_{i}^{x}+J S_{i}^{x}S_{i+1}^{x}$ as in Ref.~\cite{Zhou2017}, while the entanglement entropy of the thermal phase gradually saturates to a finite large value bounded by the maximal Page value~\cite{Page1993}. The entanglement entropy of the prethermal DTC only arrives at a small saturated value, demonstrating its non-thermalization for a steady symmetry-broken phase.

\begin{figure}[t]
  \includegraphics[height=1.95in,width=3.4in]{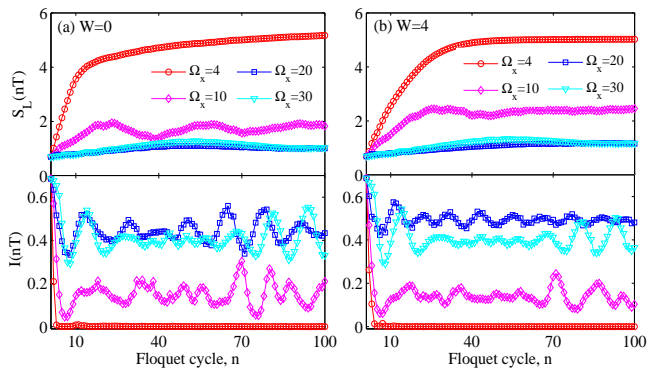}
  \caption{\label{field}(Color online) Dynamics of the half-chain entanglement entropy $S_L(t)$ and the mutual information $I(t)$ between spins on opposite ends of a length chain versus time under different $\Omega_x$ with the initial superposition states $|\psi_l(t=0)\rangle$ for disorder strength (a) $W=0$ and (b) $W=4$ respectively. The parameters $J=3,\lambda=4,L=20,\epsilon=0.038,\Omega_y\tau_2=\pi$. }
\end{figure}

Once the driving field is turned off $\Omega_x=0$, the system is dominated by the random-field Heisenberg chain $H=\sum_{i}\Delta_i S_{i}^{z}+ J(S_{i}^{x}S_{i+1}^{x}+S_{i}^{y}S_{i+1}^{y}-S_{i}^{z}S_{i+1}^{z})$, which would exhibit a MBL behavior for disorder strength $W/J\geq(W/J)_c\sim2-4$~\cite{sheng2017,vedika2017,Pal2010,Luitz2015}. For $\Omega_x=0, W/J\gg(W/J)_c$, we plot its entanglement dynamics in Fig.~\ref{mbl}(b), where the system turns into a MBL phase. In the presence of strong disorder and high driving frequency $W,\Omega\gg J$, $S_L(nT)$ has the logarithmic growth before reaching a saturated thermal value. Thus by comparing the different entanglement growth behaviors in time from Figs.~\ref{mbl}(a) with large $\Omega_x$ and~\ref{mbl}(b) without $\Omega_x$, we exclude the possibility of a Floquet-MBL protected DTC or SUP phase of our phase diagram in Fig.~\ref{phase}(b) for large driving field $\Omega_x\gg W$, demonstrating no requirement of disorder for the existence of DTC.

For $W/J\ll(W/J)_c$, we plot the behavior of the mutual information and entanglement entropy at different magnitudes $\Omega_x$ in Fig.~\ref{field}(a) and~\ref{field}(b). For large $\Omega_x$, $I(nT)$ persists to a large value close to $\log2$, and $S_L(nT)$ remains a small non-thermalized value of the order 1 at long times. When $\Omega_x$ is decreased, the prethermal condition $J/\Omega_x\ll1$ would be gradually violated~\cite{Else2017}, and the thermal dephasing effect from interaction begins to dominate. When $J\sim\Omega_x$, we find that $I(nT)$ drops dramatically down to a vanishing value, and $S_L(nT)$ quickly grows to a saturated large value bounded by the maximal thermal value, as indicated in Fig.~\ref{field}. Thus the driving field $\Omega_x$ would protect the prethermal DTC from either MBL or thermalization.

\section{Phase Transitions}\label{dynamicaltransition}

\begin{figure}[t]
  \includegraphics[height=2.5in,width=3.4in]{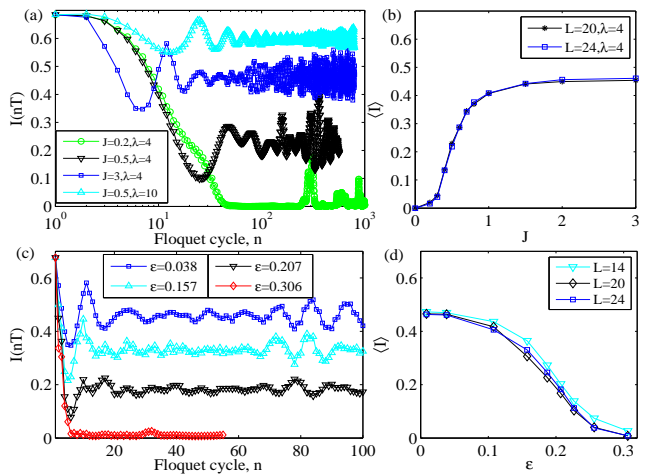}
  \caption{\label{mut}(Color online) The non-equilibrium dynamics of the mutual information between spins on opposite ends of a length chain with the initial superposition states $|\psi_l(t=0)\rangle$. (a) $I(nT)$ and (b) $\langle I\rangle$ versus $J$ at $\epsilon=0.038$. (c) $I(nT)$ and (d) $\langle I\rangle$ versus $\epsilon$ at $J=3$. The parameters $\lambda=4,\Omega_x=41.2,W=4$.}
\end{figure}

We now consider the dynamical quantum phase transition when interaction and imperfect spin flipping are tuned. The effective Hamiltonian $H_e(t)$ has a hidden emergent Ising symmetry $S^{x}\rightarrow-S^{x}$, and the dynamical transition signature from DTC to the trivial paramagnet can be obtained from the non-equilibrium dynamics of the mutual information~\cite{Keyserlingk2016,Yao2017}. For much weaker interactions $J/\Omega_x\ll0.1$, as indicated in Fig.~\ref{mut}(a), the spins are almost un-correlated and the mutual information $I(nT)$ drops from the initial value $\log2$ down to a vanishingly small time-averaged value $\langle I\rangle$ at long times. As the interaction increases $J/\Omega_x\sim0.1$, the spins are gradually correlated to each other by interaction effects, and $I(nT)$ persists at a stable large value for a long time, where the deviation from the ideal value $\log2$ may be due to imperfect flipping $\epsilon$ and traverse interaction terms. Similarly, with the initial product states $|\psi_s(t=0)\rangle$, the spin polarization $P(nT)$ also exhibits a robust $2T$-periodicity demonstrating  a prethermal phase when time $nT <t^{\ast}$. In Fig.~\ref{mut}(b), we plot the time-averaged mutual information $\langle I\rangle$ in the prethermal regime as a function of interaction strength $J$. Clearly, we see that by increasing the interaction strength or time scale $\tau_1$, the observed
DTC behavior lasts for longer time, which is also experimentally observed in Ref.~\cite{Choi2017}.

Furthermore, we consider the melting transition induced by the imperfection in the spin flipping $\epsilon$, at fixed interaction. For large $\epsilon$, the DTC behavior would be replaced by the trivial Floquet paramagnet with the rigidity of the $\omega=1/2$ peak being destroyed. As depicted in Fig.~\ref{mut}(c) and~\ref{mut}(d), the mutual information $I(t)$ shows  a steady plateau after the short thermalization time scale $2\pi/J\sim10T$, and persists to a robust nonzero value $\langle I\rangle$ in the prethermal regime. As $\epsilon$ increases, $\langle I\rangle$ gradually collapses down to zero for all the system sizes.  

\section{Summary and Discussions}\label{summary}
In summary we have presented an analysis of the prethermal Floquet DTC at long times for one dimensional systems far away from equilibrium. The prethermal Floquet DTC is characterized by the persistent time evolution of both stroboscopic spin polarization and robust mutual information between two ends, before heating up to a thermal phase.
Since the prethermal phase is mainly determined by the ratio between driving frequency and local interaction\cite{Else2017}, and does not require the protection of MBL effect from disorder, we believe the dimension effect is not important, which is consistent with the experimental observed prethermal behavior in the one-dimensional spin chain with long-range dipolar interaction~\cite{Neyenhuis2016}. It is expected that our numerical time evolution results would be relevant to experiments.

Experimentally, as $\Omega_x (\Omega_y)\gg J$, the critical short time scale determined by the Ising interaction $t_{\text{pre}}\sim2\pi/J\sim10^2T$ at $\lambda=4$ for weak interactions $J\sim0.1$, which is comparable to the current Floquet cycle. Thus the observed DTC behavor is less robust or may be in the short time regime. Once interaction $J$ is increased by an order, $t_{\text{pre}}$ is reduced by an order of magnitude and a stable prethermal DTC can be observed within the current Floquet cycle. Thus experimentally it is desirable to investigate whether the observed DTC behavior within current coherence times falls into the prethermal regime without the need of Floquet-MBL.

\begin{acknowledgements}
This research is supported by National Science Foundation Grants PREM DMR-1205734 (T.S.Z.) and DMR-1408560 (D.N.S.). We also acknowledge partial support from Princeton MRSEC DMR-1420541.
\end{acknowledgements}

\appendix

\section{Two-body reduced density matrix}\label{rdm}
In this section we describe the two-body reduced density matrix between two spins $A$ and $B$ at opposite ends. The general formulism for a two-body reduced density matrix in the eigenvector basis of $S^z$ representation is $\widehat{\rho}_{AB}=\sum_{\sigma_1\sigma_2,\sigma_3\sigma_4}\rho_{\sigma_1\sigma_2,\sigma_3\sigma_4}|\sigma_1\sigma_2\rangle\langle\sigma_3\sigma_4|$, where $\sigma=\pm\frac{1}{2}$ is the eigenvalue of $S^z$ on the eigenvector $|\sigma\rangle$. Thus in the spin basis representation $\{|\sigma\sigma'\rangle\}=\{|\uparrow\uparrow\rangle,|\uparrow\downarrow\rangle,|\downarrow\downarrow\rangle,|\downarrow\uparrow\rangle\}$ , $\widehat{\rho}_{AB}$ can be written as a $4\times4$ matrix
\begin{align}
  \widehat{\rho}_{AB}=\begin{pmatrix}
\rho_{11} & \rho_{12} & \rho_{13} & \rho_{14}\\
\rho_{21} & \rho_{22} & \rho_{23} & \rho_{24}\\
\rho_{31} & \rho_{32} & \rho_{33} & \rho_{34}\\
\rho_{41} & \rho_{42} & \rho_{43} & \rho_{44}\\
\end{pmatrix}\label{reducedmatrix}
\end{align}
The matrix elements of $\widehat{\rho}_{AB}$ are related to the two-body correlation functions
\begin{align}
  &\rho_{11}=\text{tr}\big[\widehat{\rho}_{AB}(1/2+S_{A}^{z})(1/2+S_{B}^{z})\big],\nonumber\\
  &\rho_{22}=\text{tr}\big[\widehat{\rho}_{AB}(1/2+S_{A}^{z})(1/2-S_{B}^{z})\big],\nonumber\\
  &\rho_{33}=\text{tr}\big[\widehat{\rho}_{AB}(1/2-S_{A}^{z})(1/2-S_{B}^{z})\big],\nonumber\\
  &\rho_{44}=\text{tr}\big[\widehat{\rho}_{AB}(1/2-S_{A}^{z})(1/2+S_{B}^{z})\big],\nonumber\\
  &\rho_{12}=\text{tr}\big[\widehat{\rho}_{ij}(1/2+S_{A}^{z})S_{B}^{+}\big],\nonumber\\
  &\rho_{13}=\text{tr}\big[\widehat{\rho}_{AB}S_{A}^{+}S_{B}^{+}\big],\nonumber\\
  &\rho_{23}=\text{tr}\big[\widehat{\rho}_{AB}S_{A}^{+}(1/2-S_{B}^{z})\big],\nonumber\\
  &\rho_{24}=\text{tr}\big[\widehat{\rho}_{AB}S_{A}^{+}S_{B}^{-}\big],\nonumber\\
  &\rho_{14}=\text{tr}\big[\widehat{\rho}_{AB}S_{A}^{+}(1/2+S_{B}^{z})\big],\nonumber\\
  &\rho_{34}=\text{tr}\big[\widehat{\rho}_{AB}(1/2-S_{A}^{z})S_{B}^{+}\big].\nonumber
\end{align}
Here the symmetry $\rho_{\alpha\alpha'}=\rho_{\alpha'\alpha}$. The eigenvalues $\xi_{\alpha}$ of $\widehat{\rho}_{AB}$ can be easily determined by diagonalizing Eq.~\ref{reducedmatrix}. Therefore the two-site entanglement entropy of two-body reduced density matrix is $S_{AB}=-\sum_{\alpha}\xi_{\alpha}\log\xi_{\alpha}$. Similarly, we can obtain the single-site entanglement entropy $S_{A}$ ($S_{B}$) from its one-body reduced density matrix $\widehat{\rho}_{A}$ ($\widehat{\rho}_{B}$). Finally, the mutual information between sites A and B of the spin chain $I=S_A+S_B-S_{AB}$. For the ideal long-range correlated states $|\psi_l(t=0)\rangle$, $I=\log2$ for any two sites $A,B$.


\section{Subharmonic Fourier peak}\label{sfp}
\begin{figure}[t]
  \includegraphics[height=1.45in,width=3.4in]{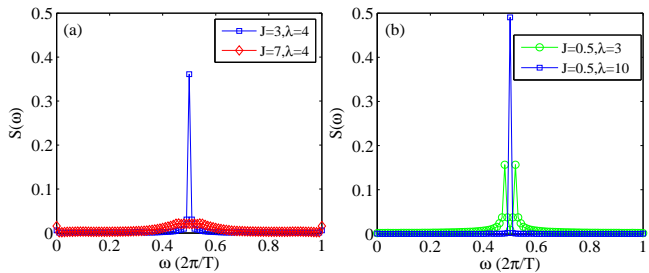}
  \caption{\label{fft}(Color online) The subharmonic Fourier response of spin polarization $P(nT)$ with the short-range correlated state $\psi_s(t=0)$ using $1\leq n\leq100$ Floquet cycles for (a) different interactions at $\lambda=4$ and (b) different time scale at $J=0.5$. The parameters $\Omega_x=41.2,W=4,\epsilon=0.038,L=24$.}
\end{figure}

\begin{figure}[b]
  \includegraphics[height=1.45in,width=3.4in]{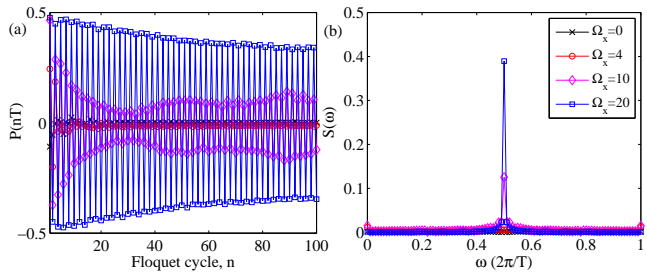}
  \caption{\label{ffthz}(Color online) (a) The spin polarization $P(nT)$ and (b) the subharmonic Fourier response with the short-range correlated state $\psi_s(t=0)$ using $1\leq n\leq100$ Floquet cycles. The parameters $J=3,\lambda=4,W=4,\epsilon=0.038,L=24$.}
\end{figure}

In this section, we study the dependence of the subharmonic Fourier peak $S(\omega)$ on the local energy scale $\Omega_x$, interaction strength $J$ and interaction time scale $\lambda=\tau_1/\tau_2$. In the absence of disorder $W=0$ and interaction $J=0$, the system is integrable, and with the short-range correlated initial states $|\psi_s(t=0)\rangle$, (i) when $\Omega_x\tau_1/(2\pi)$ is an integer, $P(t=nT)=(-1)^n\cos(n\epsilon\pi)/2$; (ii) when $\Omega_x\tau_1/(2\pi)$ is a half-integer, $P(nT)=1/2$ for even Floquet cycle $n$ and $P(nT)=-\cos(\epsilon\pi)/2$ for odd $n$. When interaction is turned on, the spins are correlated with each other. As shown in Figs.~\ref{fft}(a) and~\ref{fft}(b), for weak interaction $J\lesssim1$ and short interaction time scale $\lambda\sim1$, the subharmonic Fourier peak $S(\omega)$ exhibits two incommensurate Fourier peaks at $\omega=(1\pm\epsilon)/2$, where the phase is taken as a symmetry-unbroken phase (SUP); For very strong interactions $J\gg1$, $S(\omega)$ does not exhibit any incommensurate Fourier peaks at any $\omega$, where the phase is possibly a thermal phase; Only for an intermediate interaction strength or long interaction time scale, $S(\omega)$ exhibits one unique incommensurate Fourier peak at $\omega=1/2$, where the phase is identified as a DTC phase.

\begin{figure}[t]
  \includegraphics[height=2.0in,width=3.3in]{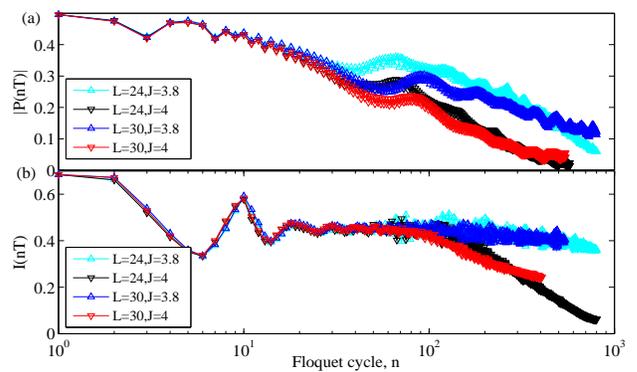}
  \caption{\label{fin}(Color online) (a) The magnitude of spin polarization $Z(nT)$ and (b) the mutual information $I(nT)$ with different chain lengths $L$. The maximal bond dimension is taken up to 400. The parameters $W=4,\lambda=4,\Omega_x=41.2,\epsilon=0.038,\Omega_y\tau_2=\pi$. }
\end{figure}

Further, given an intermediate interaction strength and interaction time scale, we plot the spin polarization $P(nT)$ and its subharmonic Fourier peak $S(\omega)$ as a function of microwave driving field $\Omega_x$ in Figs.~\ref{ffthz}(a) and~\ref{ffthz}(b). When $\Omega_x$ decreases, the spin polarization $P(nT)$ decays more quickly, the magnitude of the Fourier peak $S(\omega=1/2)$ gradually diminishes, and the steady prethermal plateaux collapses for $\Omega_x\lesssim J$. Therefore, large driving fields, acting as a local spin-pinning field, protect the spin polarization from thermalization.

\section{Finite size effects}

\begin{figure}[b]
  \includegraphics[height=2.25in,width=3.3in]{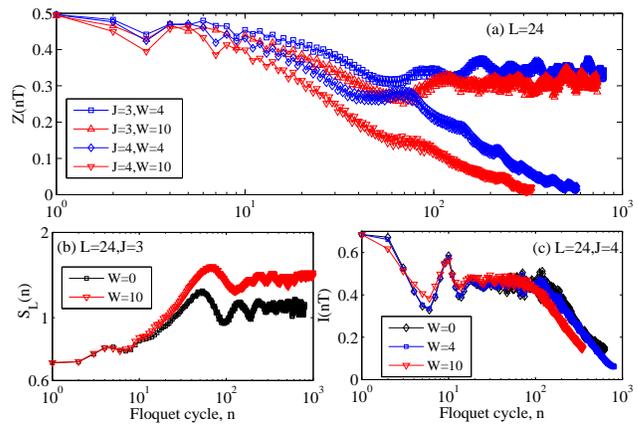}
  \caption{\label{ent}(Color online) (a) The magnitude of spin polarization $Z(nT)$, (b) the entanglement entropy $S_L(nT)$ and (c) the mutual information $I(nT)$ at different disorder strengths. The parameters $\Omega_x=41.2,\lambda=4,\epsilon=0.038,L=24$.}
\end{figure}

In this section we consider the finite system size effects on the prethermal DTC. As a prethermal DTC phase is indicated by a robust plateau of both $Z(nT)$ and $I(nT)$ before eventually heating up to a thermal phase, this picture is physically meaningful when it still persists for a finite long time in the thermodynamic size limit~\cite{Sacha2015,Russomanno2017}. The time-symmetry breaking expectation value $\langle\psi(t)|S_{i=L/2}^{x}|\psi(t)$ in the prethermal regime should fulfill two conditions: rigidity and persistence in the thermodynamic limit. In Figs.~\ref{fin}(a) and~\ref{fin}(b), we plot the stroboscopic dynamics of spin polarization and mutual information. For different chain lengths, the prethermal plateau persists as the length $L$ increases up to 30, implying its existence in the thermodynamic limit.

\section{Strong disorder}

In this section we discuss the effect of strong disorder on the robustness of the prethermal DTC, and analyze the long-time dynamics of the system for different $W$. By increasing $W$ to $W\gtrsim J$, quantitatively, the saturated value of the entanglement entropy grows up a bit, but the magnitude of stroboscopic spin polarization $Z(nT)$ decreases, as shown in Figs.~\ref{ent}(a) and~\ref{ent}(b). Further, increasing disorder strength also reduces the heating time $t^{\ast}$, as indicated in Fig.~\ref{ent}(c). It seems that for our current model, strong disorder in the $z$-direction is not favorable to the robustness of a prethermal DTC in the traverse $x$-direction.


\begin{thebibliography}{99}
\bibitem{note} Here we only consider the nearest-neighbor interaction term, due to the rapidly decaying dipolar interaction $\propto1/r^3$. This also allows us to determine a clear quantum phase diagram with reduced finite size effects.
\end{thebibliography}
\end{document}